\documentclass[useAMS,usenatbib]{rmaa}

\def\uv{uvby-\beta}

\begin{document}

\title{$\uv$ photoelectric photometry of Cepheid stars\altaffilmark{1}}

\author{J. H. Pe\~na\altaffilmark{2}, A. Arellano Ferro\altaffilmark{2}, R. Pe\~ na-Miller\altaffilmark{3},
M. Alvarez\altaffilmark{4}, Y. Rosas\altaffilmark{5}, H.
Garc\'ia\altaffilmark{6}, G. Mu\~noz\altaffilmark{7}, B.
Vargas\altaffilmark{8}, J. P. Sareyan\altaffilmark{9}, C. A. Guerrero\altaffilmark{2} \& A.
Renter\'ia\altaffilmark{2}
\medskip }

\altaffiltext{1}{Based on observations collected at the San Pedro
M\'artir Observatory, Mexico.} \altaffiltext{2}{Instituto de
Astronom\'{\i}a, Universidad Nacional Aut\'onoma de M\'exico.}
\altaffiltext{3}{Department of Mathematics, Imperial College London,
UK.} \altaffiltext{4}{Observatorio Astronomico Nacional, M\'exico}
\altaffiltext{5}{CRyA, Universidad Nacional Aut\'onoma de M\'exico}
\altaffiltext{6}{Facultad de Ciencia, UNAN-Managua,
Nicaragua.}\altaffiltext{7}{ESIME, IPN, M\'exico}
\altaffiltext{8}{Instituto de Geof\'{\i}sica, Universidad Nacional
Aut\'onoma de M\'exico} \altaffiltext{9}{Lesia, Observatoire de
Paris-Meudon and Observatoire de la Cote d'Azur, France.}

\fulladdresses{
\item J.H. Pe\~na, A. Arellano Ferro, C. A. Guerrero, \& A.
Renter\'ia: Instituto de Astronom\'{\i}a, Universidad Nacional
Aut\'onoma de M\'exico, Apdo. Postal 70-264, M\'exico D. F. CP
04510, M\'exico.
\item M. Alvarez: Instituto de Astronom\'{\i}a, Universidad
Nacional Aut\'onoma de M\'exico, Ensenada
\item J. P. Sareyan: OCA, Department Gemini, Boulevard de
l'Observatoire BP 4229, 06304 Nice Cedex 4, France.
\item R. Pe\~ na: Department of Mathematics, Imperial College London,
South Kensington Campus, London SW7 2AZ, UK}

\shortauthor{Pe\~na et al.} \shorttitle{$uvby-beta$ photometry of Cepheids}

\resumen{Presentamos fotometr\'{\i}a fotoel\'ectrica $\uv$ de 41 estrellas Cefeidas cl\'asicas. Una breve discusi\'on de los datos comparados con observaciones an\'alogas se ha llevado a cabo }

\abstract{We present time-series $\uv$ photometry of 41 classical Cepheid stars. A brief discussion of the comparison between the presented data and previous photometric data has been done. }

\keywords{Photometry - Str\"omgren photometry - stars - cepheid stars }

\maketitle

\section{INTRODUCTION}

The relevance of classical Cepheids in stellar astrophysics, both as
distance indicators and in understanding stellar structure and
pulsation, has been long acknowledged. The Str\"omgren ($\uv$) photometric system has proven to be very useful
in the determination of fundamental physical quantities such as reddening, effective temperature,
gravity and metallicity, for main
sequence (Crawford 1975), giant (Olsen 1984) and supergiant stars
(Arellano Ferro \& Parrao 1990; Arellano Ferro \& Mantegazza 1996).
Large compilations of $\uv$ photometry of Cepheids have been
published in the past (Feltz \& McNamara 1980; Eggen 1983, 1985;
Meakes et al. 1991; Arellano Ferro et al. 1998). However, while there
are some intersections in the stars studied in these works, there are
many poorly observed Cepheids. In this paper we present new
$\uv$ data for 41 Cepheids, many of which have very little or no previous
Str\"omgren photometry.

\section{Observations}

The observational data were gathered during different seasons (see
Table 1) at the San Pedro M\'artir Observatory, in Baja California,
Mexico. All the data was obtained at the 1.5 m telescope to which a
six-channel grating spectrophotometer was attached. In most of the
seasons the observational procedure was the same; each data
point reported is the average of at least five 10 s integrations and
both sets of  $uvby$ data and the narrow and wide bands that define
the $\beta$ index were observed almost simultaneously. A single
measurement of the sky with an integration time of 10 s was
subtracted from the star measurements. On each night several
standard stars were observed to carry out the transformation into
the standard system of Olsen (1983) and Crawford (1975; 1979). The
reduction procedure was done with  NABAPHOT package (Arellano Ferro
\& Parrao 1989) that corrects for atmospheric extinction, transforms
the data into the standard system and converts the sidereal time
into Heliocentric Julian Day. The standard stars were taken from
Gr\"onbech \& Olsen (1976; 1977) and Olsen (1983), but some of the
bright standard  stars were taken from the list published in
the Astronomical Nautical Almanac.

\begin{table*}[!ht]
\begin{center}
\caption[] {Log of the observing seasons.} \hspace{0.01cm}
    \label{log}
\begin{tabular}{lrccl}
\hline \hline
Epoch &   No of stars  &  Initial date  &  Final date &observers\\
      &                & year mo day     &  year mo day &         \\
\hline  \hline
1989 OctNov   & 16 &  1989 10 29 & 1989 11 07 & jhp, rpg\\
2005 MayJune      &  4 &  2005 05 28 & 2005 06 31 & jhp, rpm\\
2006 July     &  7 &  2006 07 14 & 2006 07 19 & ma, jps\\
2006 November & 12 &  2006 11 01 & 2006 11 13 & ma, lpl, jps, yr\\
2006 December &  8 &  2006 12 09 & 2006 12 11 & jhp, jps, hg\\
2007 MarchApril    &  8 &  2007 03 30 & 2007 04 03 & jhp, gm, bv\\
2007 October  & 12 &  2007 10 05 & 2007 10 26 & jhp, jps, cg\\
2008 October  &  8 &  2008 10 08 & 2008 10 14 & ma, jps\\
2008 December &  9 &  2008 12 09 & 2008 12 14 & jhp, pz, vha\\
2009 June     &  9 &  2009 06 24 & 2009 06 26 & jhp, hg, arl\\
\hline \hline
\end{tabular}
\end{center}
jhp, J.H. Pe\~na; rpg, R. Peniche; rpm, R. Pe\~na Miller; jps, J. P. Sareyan; ma, M. Alvarez; yr, Y. Rosas; lpl, L. Parrao;
hg, H. Garcia; gm, G. Mu\~noz; bv, B. Vargas; cg, C. Guerrero; pz, P. Zasche; vah, V. H. Alvarado and arl, A. Renteria 
\end{table*}\

The transformation equations used in the work have the following forms:

 V = A + B $(b-y)$(inst) + $y$ (inst)\\
 $(b-y)$ (std) = C + D $(b-y)$(inst)\\
 $m_1$(std) = E + F $m_1$(inst) + J $(b-y)$(inst)\\
 $c_1$(std) = G + H $c_1$(inst) + I $(b-y)$(inst) \\
 $\beta$ (std) = K + L $\beta$ (inst). \\

\noindent

\begin{table*}[!t]
 \caption{Mean values  and standard deviations $<\sigma>$ of the transformation coefficients
 obtained for the October, 2008 season}
\begin{center}
  \begin{tabular}{cccccccc}
\hline\hline
season          &  B      &   D    &   F    &   J    &  H     &   I   &   L\\
\hline 2008     & 0.884  & 0.996 & 1.027 & 0.013 & 1.007  & 0.062 & -1.319\\
\hline $<\sigma>$ &  0.026 & 0.015 & 0.081 & 0.031 & 0.054  & 0.074 &  0.065\\
\hline\hline
\end{tabular}
\end{center}
\label{trancoef}
\end{table*}

In Fig. 1 the transformations between the instrumental and the
standard values for a group of standard stars for the night of
October 11, 2008 are illustrated. Table 2 presents the
values of the slopes and color term coefficients averaged for seven
nights from the 2008 season. Standard deviations for each coefficient are at the bottom of this table. Except for the May 2005 season,
which was devoted entirely to the data acquisition of Cepheid stars,
most seasons were planned for the observations of short period
variable stars and hence few data points of Cepheid and standard
stars were obtained on each night. Nevertheless, some seasons were
long enough to obtain data strings suitable for the long periods of
some Cepheid stars.

\subsection{Photometric uncertainties}
Individual uncertainties were determined by calculating the standard
deviations of the fluxes in each filter for each star. It is obvious
that the brighter stars were more accurately observed than the
fainter ones. However, the faint stars were observed long enough to
secure sufficient photon counting to get high S/N ratios.
Representative values of the photon counting errors $N/\sqrt{(N)}$
derived from the measurements on the night of  October 11, 2008 for
Cepheids with magnitudes $V$ from 5.2 to 11.6 are listed in Table 3.
The standard star BS 1430 was included for comparison. In view of
the results, errors associated with photon counting appear
negligible.

Season errors were evaluated through the differences (calculated
minus reported) of the magnitude and colors for the standard stars.
Ten to fifteen standard stars were observed on each night. Emphasis
is made on the large range in the magnitude and color values of the
standard stars (see Fig. 1). We present the standard deviations of the mean values of these
differences $<\delta$($V$, $(b-y)$, $m_1$, $c_1$)$>$ = (0.012, 0.005, 0.007, 0.018)
for the October 2008 season.

\begin{table*}[!t]
 \caption{Photon counting errors on the night of $Oct11^{th}, 2008$ }
\begin{center}
  \begin{tabular}{crcccccc}
  \toprule
ID    & V & $u$  &   $b$  &    $v$ &  $y$  &N\\
  \hline\hline
BS 1430&5.4&0.0004&0.0003&0.0002&0.0003&4\\
RT AUR&5.2&0.0005&0.0003&0.0002&0.0003&5\\
SZ TAU&6.5&0.0009&0.0006&0.0004&0.0004&6\\
ST TAU&8.5&0.0023&0.0014&0.0010&0.0010&6\\
SY AUR&9.0&0.0040&0.0020&0.0020&0.0020&6\\
AO AUR&9.1&0.0023&0.0014&0.0010&0.0010&10\\
AN AUR&10.7&0.0070&0.0040&0.0020&0.0020&10\\
ER AUR&11.6&0.0090&0.0050&0.0030&0.0030&10\\
\hline\hline
\end{tabular}
\end{center}
\label{counts}
\end{table*}

\begin{table*}[!t]\centering
\setlength{\tabnotewidth}{\columnwidth}
  \tablecols{8}
 \setlength{\tabcolsep}{1.0\tabcolsep}
 \caption{Observed stars in $\uv$}
  \begin{tabular}{rccrcccc}
  \toprule
 Star & Epoch & Period & AF\tablenotemark{1} & Eggen & Meakes    &F\&McN\tablenotemark{2} & This paper  \\
      &       &        & 1998 & 1983  &  1991      &1980     &     \\

 \midrule
SW TAU&41687.77&1.583584&6& &7&&5\\
EU TAU&41324.22&2.10248&&&&&14\\
SZ TAU&34628.57&3.14873&6&&&28&46\\
SS SCT&35315.625&3.671253&26&13&&29&0\\
RT AUR&42361.155&3.728115&&&&&11\\
 Y AUR&37203.629&3.859485&1&&&&37\\
CM SCT&35111.32&3.916977&&&&&21\\
ST TAU&41761.963&4.034299&2&&&&42\\
 X SCT&34905.58&4.19807&&&&&16\\
VZ CYG&41705.702&4.864453&9&&&30&10\\
AS PER&41723.934&4.972516&&&&&13\\
BG LAC&35315.273&5.331908&&&&26&18\\
UY PER&44945.845&5.365106&&&&&3\\
BX SCT&27901.83&6.41133&&&&&11\\
AW PER&42709.059&6.463589&&&&23&11\\
AO AUR&42815.86&6.763006&&&&&15\\
CK SCT&40855.25&7.41522&&&&&13\\
RS ORI&42820.794&7.566881&10&21&&&25\\
VY CYG&43045.282&7.856982&2&&&&10\\
RX CAM&42766.583&7.912024&&&&&4\\
BK AUR&17377.719&8.002432&&&&&16\\
CN SCT&28670.16&9.9923&&&&&7\\
SY AUR&36843.52&10.144698&1&&&&36\\
AN AUR&36843.309&10.29056&&&&&46\\
 Y SCT&34947.2&10.341504&27&&&&5\\
 Z LAC&42827.123&10.885613&5&&&25&15\\
VX PER&43758.994&10.88904&&&&&4\\
TY SCT&37377.09&11.05302&&&&&8\\
SV PER&43839.296&11.129318&1&&&&14\\
RX AUR&39075.63&11.623515&&&&21&61\\
 Z Sct&36247.16&12.90133&27&&&&5\\
TX CYG&43794.971&14.7098&&&&&10\\
RW CAS&35575.227&14.7949&&&&&3\\
SZ CYG&43306.79&15.10965&10&&&&5\\
ER AUR&43861.3&15.69073&&&&&11\\
X CYG&43830.387&16.386332&18&&&53&13\\
RW CAM&37389.57&16.41437&&&&&4\\
YZ AUR&37431.141&18.193212&&&&&28\\
RU SCT&31174.67&19.70062&27&&&&8\\
VX CYG&43783.642&20.133407&6&&&&13\\
\bottomrule \tabnotetext{1} {Arellano Ferro et al. (1998)}
\tabnotetext{2}{Feltz \& McNamara (1980)}
\end{tabular}
\end{table*}

\begin{figure}
\begin{center}
\includegraphics[height=7.5cm=width=7.5cm]{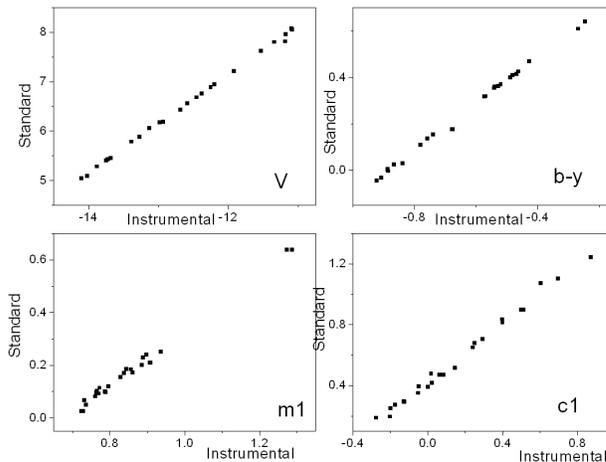}
\caption{Sample of the transformation correlations between
instrumental and standard values for a groups of standard stars for
the October, 2008 season} \label{transf}
\end{center}
\end{figure}

\renewcommand{\thefootnote}{*}

\section{Results}

A log of the observations is given in Table 4. Column 1 gives the
star name; columns two and three, the ephemerides elements employed
to calculate the phase in the light curves. These elements were
taken from the General Catalogue of Variable Stars (Samus et al.
2010) which provides correct phase light curves. Subsequent columns
in the Table report the number of observations for each star by each
one of the previously mentioned authors who provided $\uv$
photometry, namely, Feltz \& McNamara (1980), Eggen (1983, 1985),
Meakes et al. (1991) and Arellano Ferro et al. (1998). The last
column lists the number of observations the present paper provides.
The magnitudes and colors obtained in the present paper in the
standard system for our sample of Cepheids are listed in Table 5
which is available only in electronic form. A small portion of this
Table is illustrated as Table 5 in the printed version.

\begin{table*}[!t]\centering
\setlength{\tabnotewidth}{\columnwidth}
  \tablecols{13}
 \setlength{\tabcolsep}{1.0\tabcolsep}
 \caption{Sample of $\uv$ observations of classical Cepheids}
\label{compara}
  \begin{tabular}{lcccccccc}
  \toprule
 Star & $V$ & $b-y$ & $m_1$ & $c_1$ & HJD($uvby$) & $\beta$ & HJD($\beta$) & $P (days)$\\
 \midrule
SW TAU & 9.694 & 0.443 & 0.152 & 0.982 & 2447833.9053 & 2.713 &  41687.77  & 1.583584\\
SW TAU & 9.712 & 0.410 & 0.112 & 1.076 & 2447837.9903 & 2.729 &  41687.77  & 1.583584\\
SW TAU & 9.916 & 0.522 & 0.152 & 0.826 & 2454190.6381 &       &  41687.77  & 1.583584\\
SW TAU & 9.366 & 0.364 & 0.100 & 1.269 & 2454191.6274 &       &  41687.77  & 1.583584\\
SW TAU &       & 0.514 & 0.148 & 0.809 & 2454192.6355 &       &  41687.77  & 1.583584\\
\bottomrule
\end{tabular}
\end{table*}

\section{Comparison with other photometries}

The confidence level of our reported values should be judged through
the discussed uncertainties. Goodness of the quality of our data can
be demonstrated through the agreement with the previously reported
$\uv$ observations. These sources were mentioned before and the
number of data points for each star is listed in Table 4. We have
chosen four stars with a large number of observations from different
authors to demonstrate this point of agreement among these
observations. The stars considered for the comparison were X Cyg, VZ
Cyg, SW Tau and SS Tau. As can be seen in Figure 2 they are all in
perfect agreement. The comparison
between the data of Arellano Ferro et al. (1998) and those of the present paper was done for all the stars and all the observations fit
either in the light curve and color index diagrams demonstrating
once again the stability of the stars and the quality of the
observations. Figures 3 to 5 show the light curves for most of the
Cepheids in our sample. In these Figures we have included two data
sets for each star, namely Arellano Ferro et al. (1998), represented by open circles, and those of the present paper, filled
circles.

\begin{figure}
\begin{center}
\includegraphics[height=7.5cm=width=7.5cm]{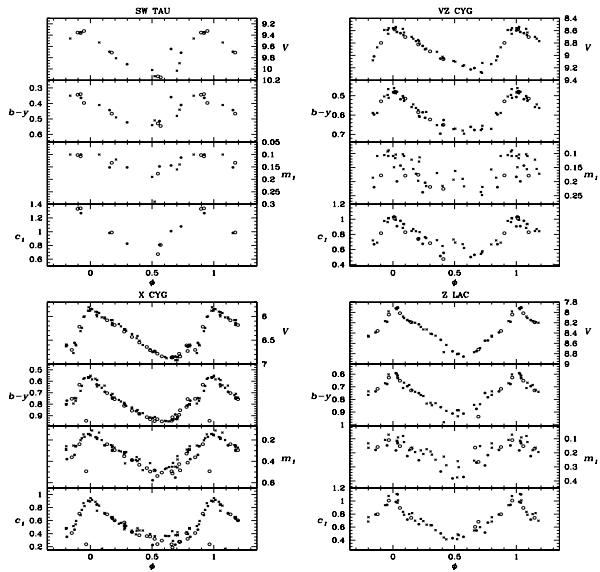}
\caption{Light curves in $\uv$ for four stars observed by several authors. The agreement between the sources shows the several data sets is good. Open circles: Arellano Ferro et al. (1998); crosses, Feltz \& McNamara (1980); filled circles, present work  }
\label{compas}
\end{center}
\end{figure}

\section{Conclusions}
New $\uv$ photoelectric photometry has been acquired and is
presented for 41 Cepheid stars. We trust that $\uv$ data, like that
presented in this work, may be useful on other fronts of Cepheid
research such as secular period changes (Szabados 1991; Arellano
Ferro 1983), metalicity (Arellano Ferro \& Mantegazza 1996),
reddening (Chulhee 2008) and other physical parameter
determinations, for instance the radii through the Baade-Wesseling
approach (Arellano Ferro \& Rosenzweig 2000). This photometry can
also be utilized to reach several other goals: to establish a
relationship with the physical properties such as the empirical
determinations developed in RR Lyrae stars (e.g. Kov\'acs and
Walker 2001 and references within) through a Fourier decomposition
of the light curves (e.g. Pe\~na et al. 2009); to determine the
metalicity photometrically from the color indexes compared directly
to the theoretical models (e.g. Meakes, Wallerstein \& Fuller
Opalko 1991); to support and improve knowledge of the chemical
enrichment gradient in the galaxy (e.g. the series of papers by
Andrievsky et al. 2004 and references within) among other topics.

\section{acknowledgments}
We would like to thank the staff of the OAN for their assistance in
securing the observations. We acknowledge the help of several
colleagues who participated in the observations: R. Peniche, L.
Parrao, V. H. Alvarado and P. Zasche. This work was partially
supported by Papiit IN110102 and IN108106-3. RPM, GM, BV, VHA and HG
thank the IA-UNAM for the opportunity to carry out the observations.
Typing was partially done by J. Orta, and proofreading by J. Miller.
C. Guzm\'an, F. Ruiz and A. D\'iaz assisted us in the computing. We
are indebted to an anonymous referee for a series of comments and
suggestions that lead to an improvement of the present paper. This
research has made use of the Simbad databases operated at CDS,
Strasbourg, France and NASA ADS Astronomy Query Form.

\begin{figure*}
\includegraphics[width=16.cm,height=18.cm]{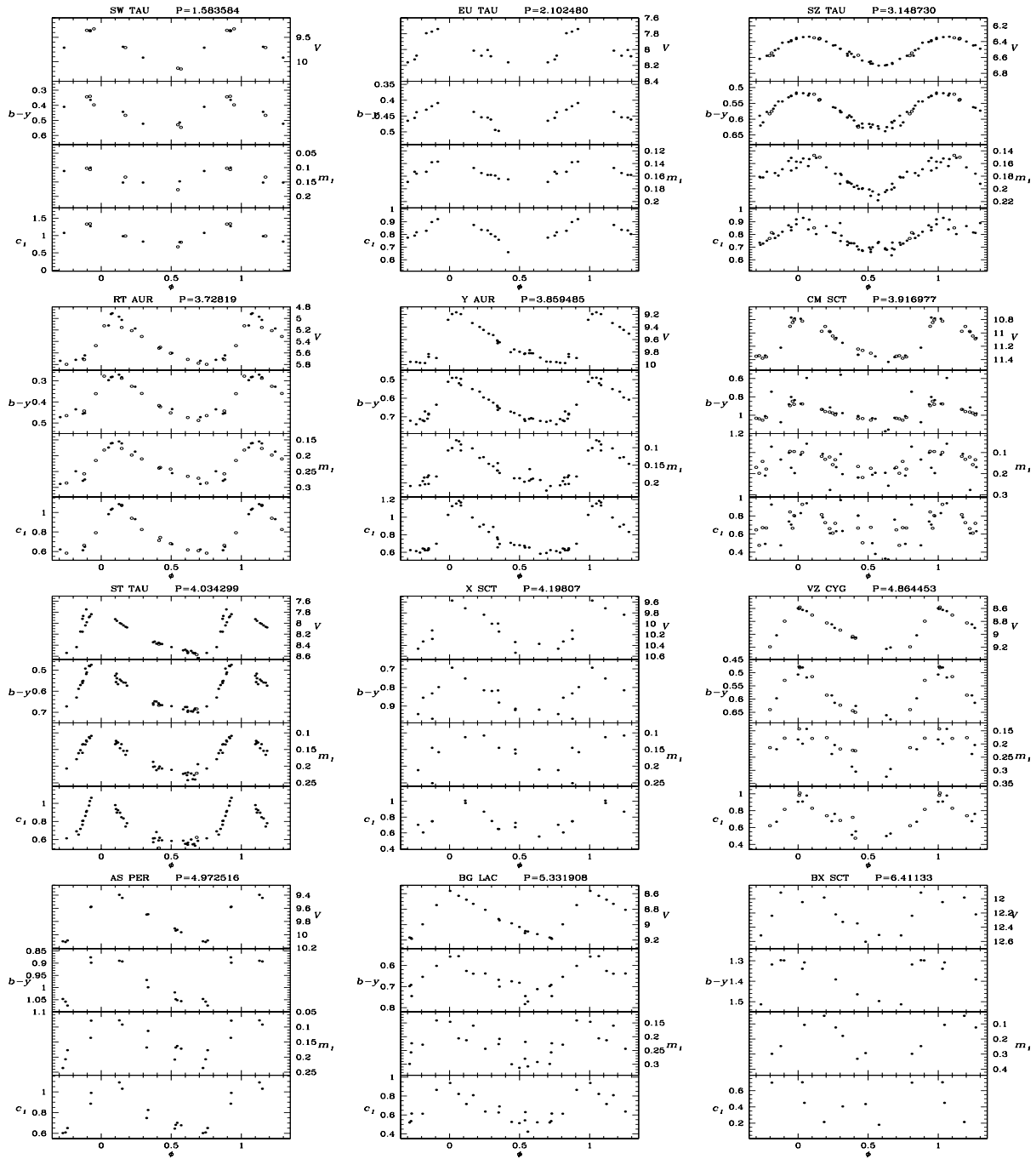}
\caption{$\uv$ Light curves of sample Cepheids}
    \label{lc}
\end{figure*}

\newpage
\begin{figure*}
\includegraphics[width=16.cm,height=18.cm]{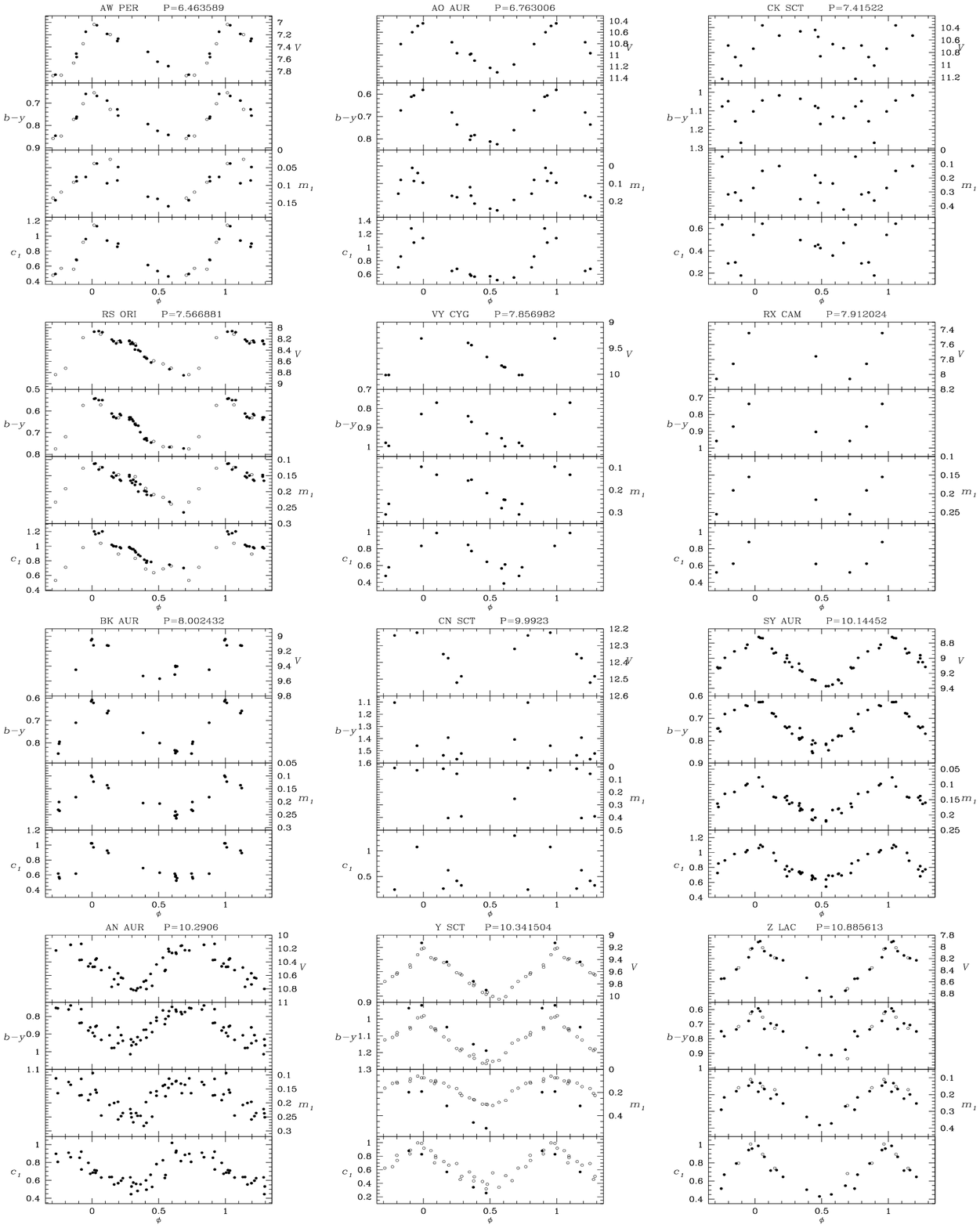}
\caption{Continue}
\end{figure*}

\newpage
\begin{figure*}
\includegraphics[width=16.cm,height=18.cm]{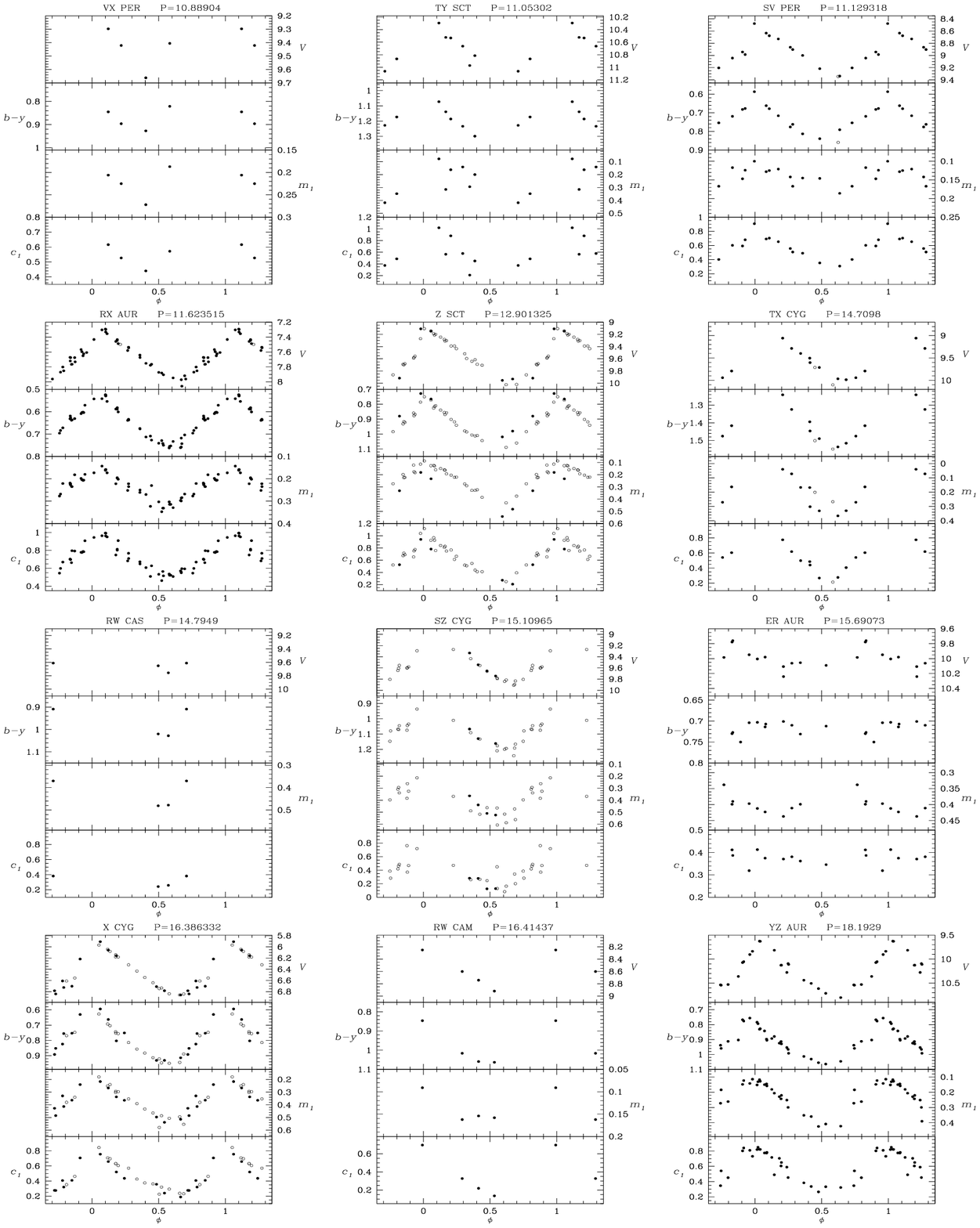}
\caption{continue}
\end{figure*}


\begin{thebibliography}


\bibitem{} Andrievsky, S. M., Luck, R. E., Martin, P. \ Lepine, J. R. D. 2004 AA 413, 159

\bibitem{} Arellano Ferro, A., 1983 ApJ 274, 755

\bibitem{} Arellano Ferro, A., Mantegazza, L., 1996, A\&A  315, 542

\bibitem{} Arellano Ferro, A. \& Parrao, L., 1989 Reporte T\'ecnico 57, IA-UNAM.

\bibitem{} Arellano Ferro, A., Rojo Arellano, E., Gonz\'alez-Bedolla, S.,  Rosenzweig, P. 1998, ApJSS 117, 167

\bibitem{} Arellano Ferro, A., Rosenzweig, P., 2000, MNRAS 315, 296

\bibitem{} Chulhee, K., 2008, ApJ 674, 1062

\bibitem{} Crawford D. L., 1975 AJ 80, 955

\bibitem{} Crawford, D. L., 1979 AJ 84, 185

\bibitem{} Eggen, O. J., 1983, AJ, 88, 998

\bibitem{} Eggen, O. J., 1985 AJ 90, 1297

\bibitem{} Feltz, K. A. \& McNamara, D. H., 1980 PASP 92, 609

\bibitem{} Gr\"onbech, B. \& Olsen, E. H., 1976 AAS 26, 155

\bibitem{} Gr\"onbech, B. \& Olsen, E. H., 1977 AAS 27, 443

\bibitem{} Kov\'acs, G. \& Walker, A. R., 2001 A\&A 371, 579

\bibitem{} Meakes, M., Wallerstein, G., \& Fuller Opalko, J., 1991, AJ 101, 1795

\bibitem{} Olsen, E. H., 1983 A\&AS 54, 55

\bibitem{} Olsen, E. H., 1984 A\&AS 57, 443

\bibitem{}Pe\~na, J. H., Arellano Ferro,  A.,  Pe\~na Miller, R., Sareyan, J. P. \& Alvarez, M.,  2009 RevMexAA 45, 191

\bibitem{} Samus, N. N. et al. 2009 General Catalogue of Variable Stars. Sternberg Astronomical Institute, Moscow, Russia

\bibitem{} Szabados, L., 1991 Mitt. Sternw. ung. Akad. Wiss, Budapest No 96

\end{thebibliography}
\end{document}